\documentclass{article}
\usepackage{spconf,amsmath,graphicx}
\usepackage[utf8]{inputenc}
\usepackage{amsmath, amssymb}
\usepackage[printonlyused,nolist]{acronym}
\usepackage{xcolor}
\usepackage{algorithm}
\usepackage{algpseudocode}
\usepackage{etoolbox}
\usepackage{multirow}
\usepackage{tikz}
\usepackage{subcaption}
\usepackage{url}
\usepackage{graphicx}
\usepackage{tikz}
\usepackage{pgfplots}
\usepackage{optidef}
\usepackage{pgf}
\usepackage{algorithm}
\usepackage{algpseudocode}
\usetikzlibrary{backgrounds,arrows}
\usepackage{LMSnodeshapes}
\usepackage{booktabs}
\pgfplotsset{compat=1.15}

\usetikzlibrary{calc}

\DeclareMathOperator*{\bw}{\boldsymbol{w}}
\DeclareMathOperator*{\bn}{\boldsymbol{n}}
\DeclareMathOperator*{\bd}{\boldsymbol{d}}
\DeclareMathOperator*{\bx}{\boldsymbol{x}}
\DeclareMathOperator*{\X}{\boldsymbol{X}}

\DeclareMathOperator*{\bp}{\boldsymbol{p}}
\DeclareMathOperator*{\bu}{\boldsymbol{u}}
\DeclareMathOperator*{\be}{\boldsymbol{e}}
\DeclareMathOperator*{\C}{\boldsymbol{C}}
\DeclareMathOperator*{\by}{\boldsymbol{y}}

\DeclareMathOperator*{\herm}{\text{H}}

\DeclareMathOperator*{\bPsi}{\boldsymbol{\Psi}}

\begin{acronym}
	\acro{OSASI}{Online Supervised Acoustic System Identification}
	\acro{SNR}{signal-to-noise ratio}
	\acro{AEC}{acoustic echo cancellation}
	\acro{PF}{post filter}
	\acro{LMS}{least mean square}
	\acro{NLMS}{normalized LMS}
	\acro{RLS}{Recursive Least Squares}
	\acro{DFT}{discrete Fourier transform}					
	\acro{DTFT}{discrete time Fourier transform}
	\acro{MAP}{Maximum A Posteriori}
	\acro{MMSE}{Minimum Mean Square Error}
	\acro{STFT}{Short-Time Fourier Transform}
	\acro{PSD}{power spectral density}
	\acro{ERLE}{echo return loss enhancement}
	\acro{RIR}{room impulse response}
	\acro{FIR}{finite impulse response}							
	\acro{IR}{impulse response}
	\acro{pdf}{probability density function}
	\acro{VSSS}{variable step-size selection}
	\acro{VSS}{variable step-size}
	\acro{FD}{frequency-domain}
	\acro{TD}{time-domain}
	\acro{MM}{Minorize-Maximization}
	\acro{KF}{Kalman filter}
	\acro{DNN-FDAF}{deep neural network-controlled frequency-domain adaptive filter} % rd
	\acro{FDAF}{frequency-domain adaptive filter}
	\acro{DNN}{deep neural network}
	\acro{AIR}{acoustic impulse response}
	\acro{ATF}{acoustic transfer function}
	\acro{FR}{frequency response}
	\acro{NMF}{nonnegative matrix factorization}
	\acro{MSE}{mean squared error}
	\acro{GRU}{gated recurrent unit}
\acro{NESD}{normalized Euclidean system distance}
\end{acronym}

\pgfdeclarelayer{foreground}
\pgfdeclarelayer{background}
\pgfsetlayers{background,main,foreground}

\newtoggle{long}
\toggletrue{long}
\newtoggle{blind}
\toggletrue{blind}
\newcommand{\commentTHa}[1]{\textcolor{black}{#1}}
\newcommand{\commentTHb}[1]{\textcolor{black}{#1}}

\newcommand{\commentTHe}[1]{\textcolor{black}{#1}}

\newcommand{\commentTHd}[1]{\textcolor{black}{#1}}
\newcommand{\commentTHf}[1]{\textcolor{black}{#1}}
\newcommand{\commentTHg}[1]{\textcolor{black}{#1}}

\newcommand{\commentTHj}[1]{\textcolor{black}{#1}}			% this was the one including review comments

\title{End-to-End Deep Learning-Based Adaptation Control for Frequency-Domain Adaptive System Identification}
\name{Thomas Haubner, Andreas Brendel, and Walter Kellermann\thanks{{This work was partially funded by the German Research Foundation {- 282835863 -} within the Research Unit FOR2457 Acoustic Sensor Networks.}}}
\address{Multimedia Communications and Signal Processing, Friedrich-Alexander-University Erlangen-Nürnberg,\\ Cauerstr. 7, D-91058 Erlangen, Germany, thomas.haubner@fau.de}

\begin{document}
	\ninept
	
	\maketitle

%	\setlength{\abovedisplayskip}{5.0pt plus 3.0pt minus 4.0pt}
%	\setlength{\belowdisplayskip}{5.0pt plus 3.0pt minus 4.0pt}
%	\setlength{\abovedisplayshortskip}{0.0pt plus 2.0pt}
%	\setlength{\belowdisplayshortskip}{3.0pt plus 2.0pt minus 2.0pt}
%	
%	\the\abovedisplayskip
%
%	\the\belowdisplayskip
%	
%	\the\abovedisplayshortskip
%	
%	\the\belowdisplayshortskip
	
	\begin{abstract}
		% 140 words
		We present a novel end-to-end deep learning-based adaptation control algorithm for frequency-domain adaptive system identification. The proposed method exploits a deep neural network to map observed signal features to corresponding step-sizes which control the filter adaptation. The parameters of the network are optimized in an end-to-end fashion by minimizing the average \commentTHd{normalized} system distance of the adaptive filter. This avoids the need of explicit signal power spectral density estimation as required for model-based adaptation control and further auxiliary mechanisms to deal with model inaccuracies.
		The proposed algorithm achieves fast convergence and robust steady-state performance for scenarios characterized by \commentTHf{high-level,} non-white and non-stationary \commentTHf{additive} noise signals, \commentTHf{abrupt} environment changes and additional model inaccuracies. 
	\end{abstract}
	\begin{keywords}
		System Identification, Adaptation Control, Deep Learning, \commentTHj{Step-Size Estimation, Acoustic Echo Cancellation}%Frequency-Domain Adaptive Filtering, \commentTHj{Acoustic Echo Cancellation, Step Size Estimation}
	\end{keywords}
	%
	% TODO: add keyword "step-size", "step-size estimation" or "step-size control"
	%
	\section{Introduction}
	\label{sec:intro}
	Adaptive system identification is required for many modern signal enhancement approaches, e.g., in full-duplex acoustic communication devices for the purpose of \ac{AEC} \cite{enzner_acoustic_2014}. Despite the recently increased focus on direct deep learning-based signal enhancement algorithms \cite{zhang_deep_2019,westhausen2020acoustic}, the benefit of additionally using model-based system identification has shown to be beneficial, especially when computational load should be minimized \cite{combAdFiltAndComValDPF, kfNN}.
	
	However, the benefits of physical models are only fully {exploited if \commentTHf{their} optimum} model parameters can be quickly and robustly identified. For this\commentTHf{,} gradient descent-based model parameter updates have proven to be a powerful approach \makebox{\cite{haykin_2002, diniz_adaptive_filtering}} when combined with carefully-designed adaptation control to deal with interfering signals and noise (jointly termed 'noise' in the sequel) \cite{haensler2004acoustic} and model inaccuracies, e.g., undermodeling the filter length \cite{undermodeling}.
	During the last decades a plethora of adaptation control methods have been proposed, ranging from simplistic binary stall-or-adapt approaches \makebox{\cite{gansler_double-talk_1996, Benesty_new_2000}} to sophisticated model-based step-size estimators \cite{6112248, nitsch2000frequency, benesty-vss-lms, hauemmer_kalman_nlms} and learning-based combination of step-size selection schemes \cite{breining_applying_nodate}. Most adaptation control approaches assume simplifying probabilistic signal models to estimate a time-varying step-size by minimizing the \acl{MSE} signal or a system distance between the estimated \ac{FR} and the true \ac{FR}. In particular the \commentTHd{frequency-selective} step-size inference by a diagonalized \ac{DFT}-domain \ac{KF} \cite{enzner_frequency-domain_2006} performs robustly in scenarios challenged by non-white and non-stationary \commentTHf{additive} noise signals. However, the performance of these model-based approaches depends on the validity of the model assumptions and robust estimation of the required statistics \cite{yang_frequency-domain_2017}. Especially the estimation of statistics corresponding to unobserved quantities, e.g., the noise \ac{PSD}, poses a difficult problem \cite{yang_frequency-domain_2017}. Various estimators have been proposed, e.g., \cite{malik_online_2010, franzen_improved_2019, jiang_improved_2019}, whose performance often crucially \commentTHf{depends} on the choice of additional hyperparameters  and the considered application. % and the considered application  
	Recently, the exploitation of machine learning-based noise \ac{PSD} estimators \cite{kfNN, kfNMF} has shown significant improvements relative to classical, i.e., non-trainable, estimators. Yet, these approaches still rely on simplistic random walk models to describe the temporal evolution of the \ac{FR} \cite{enzner_frequency-domain_2006} and require a sophisticated cost function design for \ac{PSD} estimation \cite{nugraha_multichannel_2016}.
	% TODO: and unclear defintion of optimum cost function for estimating signal statistics maybe cite Nughara
	%
	% TODO: add keyword "step-size", "step-size estimation" or "step-size control"
	%
	
	Thus, we introduce in this paper an end-to-end deep learning-based adaptation control {algorithm} for frequency-domain adaptive system identification which we {term} {\ac{DNN-FDAF}}. We propose to learn a mapping from observable signal features to step-sizes by a \acs{DNN} with the average \commentTHd{\ac{NESD}} of the step-size-controlled adaptive filter as a loss function. 
	% By optimizing the \acs{DNN} parameters directly w.r.t. the \commentTHd{\ac{NESD}}, we avoid the estimation of auxiliary signal statistics for model-based adaptation control whose effect on the system identification performance is unclear if the model assumptions are not fulfilled perfectly.
	%
	By optimizing the \acs{DNN} parameters directly w.r.t. the \commentTHd{\ac{NESD}}, we avoid the estimation of auxiliary signal statistics for model-based adaptation control whose effect on the system identification performance \commentTHg{depends on the validity of the assumed model properties.}
	%
	%is unclear if the model assumptions are not fulfilled perfectly.
	%
	%
	In addition, it circumvents the need for application-dependent hyperparameter tuning inherent to many baseline approaches. The proposed method combines the benefits of physically-motivated system identification approaches, e.g., generalization to unknown environments, with the {excellent} modeling capability of \acs{DNN}s for adaptation control.
	% T
	% TODO: Ich würde noch argumentieren, dass die Schrittweite so tatsächlich den Systemabstand adressiert und alle Zwischengrößen an die bestmögliche Schätzung dieser Größe angepasst werden was ein großer Vorteil gegenüber bestehenden Verfahren ist.
	% TODO: cite early approaches from Breinging !!!!!
	
	% \commentTHf{We indicate time-domain quantities by underlined symbols and use bold lowercase letters for vectors and bold uppercase letters for matrices.}
	We use bold lowercase letters for vectors and bold uppercase letters for matrices with underlined symbols indicating time-domain quantities. 
	The identity matrix and \ac{DFT} matrix of dimensions {\makebox{$D \times D$}} are denoted by $\boldsymbol{I}_D$ and {$\boldsymbol{F}_D$}, respectively, and the all-zero matrix of dimensions $D_1 \times D_2$ by $\boldsymbol{0}_{D_1 \times D_2}$. Furthermore, we introduce the $\text{diag}(\cdot)$ operator which generates a diagonal matrix from its vector-valued argument and indicate its $m$th element by $\left[ \cdot \right]_{mm}$. The transposition and Hermitian transposition of a matrix are represented by $(\cdot)^{\text{T}}$ and $(\cdot)^{\herm}$ , respectively. Finally, we use the Euclidean norm $||\cdot||_2$ and the expectation operator $\mathbb{E}[\cdot]$.
	
	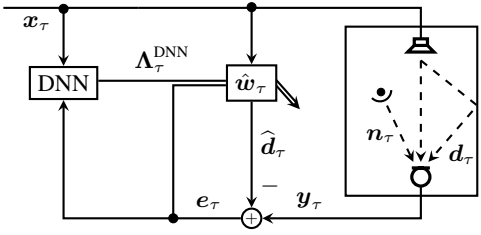
\begin{figure}[t!] % TODO: maybe directly 
		%\vspace*{-.17cm}
		\centering
			
	\begin{tikzpicture}[node distance=1.5cm, >=stealth, cross/.style={path picture={ 
			\draw[black]
			(path picture bounding box.south east) -- (path picture bounding box.north west) (path picture bounding box.south west) -- (path picture bounding box.north east);
	}}]
	\tikzset{loudspeaker style/.style={%
			draw,very thick,shape=loudspeaker,minimum size=5pt
	}}
	\tikzset{microphone style/.style={%
			draw,very thick,shape=microphone,minimum size=.5pt, inner sep=2.0pt
	}}
	
	\begin{pgfonlayer}{foreground}
	\node (sigInX) at (0,0) {};
	\node [right of=sigInX, inner sep=0] (spltX) {};
	\node [below of=spltX, rectangle, draw, thick, node distance=1.0cm, fill=white] (pbkf) {$\hat{\boldsymbol{w}}_\tau$};
	\node [below of=pbkf, circle, draw, inner sep=2.6, thick, node distance=1.8cm] (subtraction) {};
	\node (addSymb) at (subtraction) {\tiny$+$};

	\node[loudspeaker style,rotate=-90] (speaker) at (3.75,-.5) {};
	\node[microphone style,rotate=90] (mic) at (3.75,-2.25) {};
	
	\draw [thick] ($(sigInX)$) -| (speaker.west) node [below, pos=-.18] {{${\boldsymbol{x}}_{\tau}$}};
	%\draw [->, thick] (sigInX) -| (pbkf);
	\draw [->, thick] (pbkf) -- (subtraction.north) node [pos=.4, right] {$\widehat{{\boldsymbol{d}}}_{\tau}$} node [pos=.8, right] {$-$};
	\draw [->, thick] (mic.west) |- (subtraction.east) node [pos=.85, above] {${\boldsymbol{y}}_{\tau}$};
	\draw [black,fill=black] ($(pbkf.north) + (0, 0.76)$) circle (0.6mm);
	\draw [->, thick] ($(pbkf.north)+(0,.72)$) -- (pbkf.north);

	\draw [thick] (2.75,-2.5) rectangle (4.5,-.25);

	\draw [thick] ($(sigInX)+(-1.8,0)$) -- ($(sigInX)$);
	
	\draw [thick, dashed, ->] ($(speaker)-(.0,.2)$) -- ($(mic)+(.0,.2)$);
	\draw [thick, dashed, ->] ($(speaker)-(.0,.2)$) -- ($(speaker)+(.75,-.8)$) -- ($(mic)+(.1,.2)$);
	
	\node [] (d) at ($(mic)+(.55,.3)$) {${\boldsymbol{d}}_\tau$}; 
	
	% \node [] (s) at ($(mic)+(-.55,.3)$) {$\boldsymbol{d}_\tau$};
	\draw [fill] (3.22,-1.12) circle (.05);
	\draw [thick] (3.1,-1.2) arc(225:360:.15);
	\node [] (s) at ($(3.22,-1.70)$) {${\boldsymbol{n}}_\tau$};
	\draw [dashed, ->, thick] ($(3.3,-1.3)+(0,0)$) -- ($(mic)+(-.1,.2)$);

%	\draw [fill] (2.8,-2.2) circle (.05);
%	\draw [thick] (2.9,-2.3) arc(-45:45:.15);
%	\draw [dashed, ->, thick] ($(3.0,-2.2)+(0,0)$) -- ($(mic)+(-.2,.10)$);
%	\node [] (s) at ($(3.32,-2.45)$) {${\boldsymbol{n}}_\tau$};

	% include adaptation control
	\node [left of=pbkf, rectangle, draw, thick, node distance=2.5cm] (adControl) {DNN};
	
	\draw [black,fill=black] ($(adControl.north) + (0, 0.76)$) circle (.6mm);
	\draw [thick, ->]  ($(adControl.north) + (0, 0.76)$) -- (adControl.north);
	
	\draw [thick, ->] (subtraction.west) -| (adControl.south) node [pos=.09, above] {${\be}_\tau$};
	\draw [thick] ($(pbkf.west)+(-.7,-1.8)$) |- ($(pbkf.west)+(.0,-.03)$);
		\draw [black,fill=black] ($(pbkf.west)+(-.7,-1.8)$) circle (.6mm);
	
	\draw [thick] ($(adControl.east)+(0,+.03)$) -- ($(pbkf.west)+(-.0,+.03)$) node [above, pos=.5, yshift=0] {$\boldsymbol{\Lambda}_\tau^{\text{DNN}}$};
	% \draw [thick] ($(adControl.east)+(0,-.0)$) -- ($(pbkf.west)+(-.0,-.0)$) node [below, pos=.3, yshift=0] {$\boldsymbol{M}_\tau$};
	
	% \draw [thick] ($(adControl.east)+(0,+.075)$) -| ($(pbkf.west)+(-.2,+.075)$) node [above, pos=.3, yshift=-2] {$\boldsymbol{\Lambda}_\tau$};
	% \node [right of=adControl, circle, cross, draw, inner sep=2.4, thick, xshift=-.2cm, yshift=-.15cm] (mulFac) {};
	% \draw [thick] (mulFac.east) -| ($(pbkf.west)+(-.2,.025)$);
	% \draw [thick] ($(adControl.east)+(0,-.15)$) -- (mulFac.west) node [below, pos=.5, yshift=-0] {$\boldsymbol{M}_\tau$};;
	% \draw [thick] ($(mulFac.south)+(0,-1.62)$) -- (mulFac.south);
	% \draw [thick, double] ($(pbkf.west)+(-.2,.05)$) -- ($(pbkf.west)+(0,.05)$);
	
	\end{pgfonlayer}

  \begin{pgfonlayer}{background}
		\draw [double, thick, ->] ($(pbkf.east)+(-.1,+.1)$) -- ($(pbkf.east)+(.3,-.35)$);	
	\end{pgfonlayer}

	\end{tikzpicture}
		\caption{{Block diagram of the proposed \ac{DNN-FDAF} \commentTHd{algorithm} for online system identification.}}	
		% \caption{{Block diagram of the proposed} \ac{DNN-FDAF} adaptation control algorithm.}	
		%\vspace*{-1.5cm}
		\label{fig:alg_overview}
	\end{figure}
	
	\section{Adaptive System Identification}
	\label{sec:adSysId}
	%
	% source to receiver
	{In the following, the online identification of an acoustic model describing the multi-path propagation from a source, e.g., a loudspeaker, to a microphone as shown in Fig.~\ref{fig:alg_overview} is considered. We model a block of noisy \commentTHj{time-domain} microphone observations at block index $\tau$}
	\begin{equation}
		\underline{\by}_\tau = \begin{pmatrix}
			\underline{y}_{\tau R - R + 1}, \underline{y}_{\tau R - R + 2}, \dots, \underline{y}_{\tau R}
		\end{pmatrix}^{{\text{T}}} \in \mathbb{R}^{R} 
		\label{eq:td_block_y}
	\end{equation}
	as a linear superposition
	\begin{equation}
		{\underline{\boldsymbol{y}}}_\tau = {\underline{\boldsymbol{d}}}_\tau + {\underline{\boldsymbol{n}}}_\tau  \in \mathbb{R}^{R}
		\label{eq:timeDomObsEq}
	\end{equation}
	{of a noise-free observation component $\underline{{\bd}}_\tau$ and a noise component $\underline{{\bn}}_\tau$.}
	{The} noise-free observation {component} $\underline{\bd}_{\tau}$ is described by a linear convolution of {an observable input} signal block
	\begin{equation}
		\underline{\bx}_\tau = \begin{pmatrix}
			\underline{x}_{\tau R - M + 1}, \underline{x}_{\tau R - M + 2}, \dots, \underline{x}_{\tau R}
		\end{pmatrix}^{{\text{T}}} \in \mathbb{R}^{M} 
		\label{eq:timeDomInSig}
	\end{equation}
	with a {\ac{FIR}} filter \makebox{$\underline{\bw}_\tau \in \mathbb{R}^L$} of length \makebox{$L=M-R$} {modeling the multi-path propagation}. 
	The linear convolution can efficiently be implemented by overlap-save processing in the \ac{DFT} domain
	\begin{equation}
		\underline{\bd}_{\tau} = \boldsymbol{Q}_1^{{\text{T}}} \boldsymbol{F}_M^{-1} \boldsymbol{X}_{\tau} {\bw}_{\tau} \in \mathbb{R}^{R}\commentTHf{,}
		\label{eq:timeDomObsEqWithDFT}
	\end{equation}
	with the {\ac{FR}} \makebox{${\bw}_{\tau} = \boldsymbol{F}_M \boldsymbol{Q}_2 \underline{\bw}_\tau \in \mathbb{C}^M$}, the \ac{DFT}-domain {input} signal matrix \makebox{$\X_{\tau} = \text{diag} \left({\boldsymbol{x}}_{\tau} \right) = \text{diag} \left( \boldsymbol{F}_M \underline{\boldsymbol{x}}_{\tau} \right) \in \mathbb{C}^{M \times M}$} and the zero-padding matrix \makebox{$\boldsymbol{Q}_2^{\text{T}}= \begin{pmatrix}\boldsymbol{I}_{M-R} & \boldsymbol{0}_{M-R \times R }\end{pmatrix}$}. Note that \makebox{$\boldsymbol{Q}_1^{\text{T}} = \begin{pmatrix}\boldsymbol{0}_{R \times M-R} & \boldsymbol{I}_R\end{pmatrix}$} ensures a linear convolution by constraining the inverse \ac{DFT} of the product {$\boldsymbol{X}_{\tau} {\bw}_{\tau}$}.
	By inserting the propagation model \eqref{eq:timeDomObsEqWithDFT} into the signal model \eqref{eq:timeDomObsEq} and pre-multiplying with the transformation matrix $\boldsymbol{F}_M \boldsymbol{Q}_1$, we obtain the \ac{DFT}-domain observation model
	\begin{equation}
		{\by}_{\tau} = {\C}_{\tau}  {\bw}_{\tau} + {\bn}_{\tau} \in \mathbb{C}^M{,}
		\label{eq:fd_obs_eq}
	\end{equation}
	with \makebox{$\boldsymbol{C}_\tau = \boldsymbol{F}_M \boldsymbol{Q}_1 \boldsymbol{Q}_1^{\text{T}} \boldsymbol{F}_M^{-1} \X_{\tau}$} being the overlap-save constrained {input} signal matrix and the \ac{DFT}-domain microphone and noise blocks \makebox{$\by_{\tau} = \boldsymbol{F}_M \boldsymbol{Q}_1 \underline{\by}_{\tau} $} and \makebox{$\bn_{\tau} = \boldsymbol{F}_M \boldsymbol{Q}_1 \underline{\bn}_{\tau}$}, respectively.

	The {unknown \ac{FR}} ${\bw}_\tau$ is usually estimated by iteratively applying the update rule \cite{ferrara_fast_1980}
	\begin{align}
		{\be}_\tau & = {\by}_\tau - \widehat{\bd}_\tau = {\by}_\tau - \boldsymbol{C}_\tau \hat{\bw}_{\tau - 1} \label{eq:errComp} \\
		\hat{\bw}_{\tau}  & = \hat{\bw}_{\tau - 1} + \boldsymbol{Q}_3 \boldsymbol{\Lambda}_{\tau} \boldsymbol{X}_\tau^{\herm} {\be}_\tau \label{eq:filtUpd}
	\end{align}
	%
	% TODO: maybe shortly discuss that the step-size matrix also changes the orientation of the gradient update?
	%
	which represents a block-based frequency-domain implementation of the {\ac{LMS}} algorithm \cite{widrow_b_adaptive_1960} and is often termed \acs{FDAF} \cite{haykin_2002, shynk}. Here, the preceding estimate $\hat{\bw}_{\tau - 1}$ is updated by a multiplication of the stochastic gradient $\boldsymbol{X}_\tau^{\herm} {\be}_\tau$, including the prior error $\be_{\tau}$, with a diagonal step-size matrix {\makebox{$\boldsymbol{\Lambda}_{\tau} \in \mathbb{R}^{M \times M}$}}. Note that the gradient projection matrix \makebox{$\boldsymbol{Q}_3 = \boldsymbol{F}_M \boldsymbol{Q}_2 \boldsymbol{Q}_2^{\text{T}} \boldsymbol{F}_M^{-1}$} ensures {that the \ac{FR} estimate $\hat{\bw}_\tau$ corresponds to a {zero-padded} time-domain \ac{FIR} filter \makebox{$\hat{\underline{\bw}}_\tau = \boldsymbol{Q}_2^{\text{T}} \boldsymbol{F}_M^{-1} \hat{{\bw}}_\tau $}.}
	
	\section{Adaptation Control}
	\label{sec:adControl}
	%
	%The convergence rate, steady-state performance and noise-robustness of the adaptive system identification algorithm described in Sec.~\ref{sec:adSysId} decisively depends on an accurate choice of the \commentTHd{diagonal} step-size matrix $\boldsymbol{\Lambda}_{\tau}$. In the following, we will introduce a \acs{DNN}-based {method} which infers the step-size matrix $\boldsymbol{\Lambda}_{\tau}$ from the observed signals $\boldsymbol{x}_\tau$ and ${\be}_\tau$ as shown in Fig.~\ref{fig:alg_overview}.
	%
	The convergence rate, steady-state performance and noise-robustness of the adaptive system identification algorithm described in Sec.~\ref{sec:adSysId} decisively depends on an accurate choice of the \commentTHd{frequency-dependent step-sizes contained in} $\boldsymbol{\Lambda}_{\tau}$. In the following, we will introduce a \acs{DNN}-based {method} which infers the step-size matrix $\boldsymbol{\Lambda}_{\tau}$ \commentTHf{directly} from the observed signals $\boldsymbol{x}_\tau$ and ${\be}_\tau$ as shown in Fig.~\ref{fig:alg_overview}.
	%
	% In the following, we will introduce a \acs{DNN}-based {method} which infers the step-size matrix $\boldsymbol{\Lambda}_{\tau}$ from the observed signals $\boldsymbol{x}_\tau$ and ${\be}_\tau$ as shown in Fig.~\ref{fig:alg_overview}.
	% 

	\subsection{{Model-Based} Adaptation Control}
	\label{sec:impSigStatForAdCon}
	We start by discussing state-of-the-art model-based adaptation control which will serve as a motivation for the proposed method in Sec.~\ref{sec:dnn_based_step_size_ad}.
	The vast majority of these approaches {suggests a computation of the step-size matrix}
	\begin{equation}
		{\boldsymbol{\Lambda}_{\tau}^{\text{MB}}} = f_{\text{MB}} \left( {\bPsi}_{\tau}^{\text{XX}}, {\bPsi}_{\tau}^{\text{NN}}, {\bPsi}_{\tau}^{\Delta\text{W}\Delta\text{W}}\right)
		\label{eq:mapping_stepSize_mb}
	\end{equation}
	from the input signal \ac{PSD} matrix \makebox{$ \bPsi_{\tau}^{\text{XX}} = \mathbb{E} \left[  {\bx}_\tau {\bx}_\tau^{\herm} \right]$}, the noise signal \ac{PSD} {matrix} \makebox{$ \bPsi_{\tau}^{\text{NN}} = \mathbb{E} \left[  {\bn}_\tau {\bn}_\tau^{\herm} \right]$} and \commentTHg{a correlation matrix representing the \ac{FR} estimation uncertainty}
	 %the uncertainty of the \ac{FR} estimate
	 \makebox{$\bPsi_{\tau}^{\Delta\text{W}\Delta\text{W}} = \mathbb{E} \left[ \Delta \bw_{\tau} \Delta \bw_{\tau}^{\herm} \right]$} with \makebox{$\Delta \bw_{\tau} = \hat{\bw}_{\tau} - \bw_{\tau} $}.
	% {with $f_{\boldsymbol{\Lambda}_{\tau}}$ being different mappings.}
	Prominent examples are the classical \acs{FDAF} update \cite{haykin_2002, shynk}
	\begin{equation}
		\left[\boldsymbol{\Lambda}_{\tau}^{\text{FDAF}} \right]_{mm} = \frac{\mu_{\text{FDAF}}}{ \left[\bPsi_{\tau}^{\text{XX}} \right]_{mm} }
		\label{eq:fdaf_stepSize}
	\end{equation}
	with $\mu_{\text{FDAF}} ~\commentTHf{> 0}$ being a hyperparameter, and the diagonalized \commentTHd{\ac{DFT}-domain} \ac{KF} update \cite{enzner_frequency-domain_2006, malik_online_2010, franzen_improved_2019}
	\begin{equation}
		\left[\boldsymbol{\Lambda}_{\tau}^{\text{KF}} \right]_{mm} = \frac{ \left[ {\bPsi}_{\tau}^{\Delta\text{W}\Delta\text{W}} \right]_{mm} }{  \left[  {\bx}_\tau {\bx}_\tau^{\herm} \right]_{mm} \left[\bPsi_{\tau}^{\Delta\text{W}\Delta\text{W}} \right]_{mm} + \frac{M}{R}\left[ \bPsi_{\tau}^{\text{NN}} \right]_{mm} }\commentTHg{,}
		\label{eq:stepSizeOptKf}
	\end{equation}
	\commentTHg{where the} latter can be considered as a noise-robust state-of-the-art \commentTHd{approach.} 
	%Note that the factor $\frac{M}{R}$ in Eq.~\eqref{eq:stepSizeOptKf} is a normalization factor which results {from zero-padding $\underline{\bn}_{\tau}$ (cf.~Eq.~\eqref{eq:timeDomObsEq}) before applying the \ac{DFT} to obtain Eq.~\eqref{eq:fd_obs_eq}} \cite{enzner_frequency-domain_2006}. 
	\commentTHd{However, the system identification} performance of these model-based approaches crucially depends on a robust estimation of the respective signal statistics (cf.~Sec.~\ref{sec:intro}). While the estimation of the {input} signal \ac{PSD} {matrix} ${\bPsi}_{\tau}^{\text{XX}}$ is {straightforward} due to the \commentTHf{observability} of ${\bx}_{\tau}$, the estimation of statistics corresponding to unobserved quantities, e.g., ${\bPsi}_{\tau}^{\text{NN}}$ and ${\bPsi}_{\tau}^{\Delta\text{W}\Delta\text{W}}$, {is still \commentTHf{a not sufficiently solved} problem and} has been explored extensively \cite{malik_online_2010, franzen_improved_2019}.
	
	\subsection{Deep Neural Network-based Adaptation Control}
	\label{sec:dnn_based_step_size_ad}
	
	We suggest to replace the model-based mapping of signal statistics to step-size matrices (cf.~Eq.~\eqref{eq:mapping_stepSize_mb}) by a learned mapping {$f_{\text{ML}} $} of the observable {input} signal sequence ${\bx}_{1}, \dots, {\bx}_{\tau}  $ and prior error signal sequence $ {\be}_{1}, \dots, {\be}_{\tau} $:
	\begin{equation}
		\boldsymbol{\Lambda}_{\tau}^{\text{DNN}} = f_{\text{ML}} \left( {\bx}_{1}, {\be}_{1}, \dots, {\bx}_{\tau}, {\be}_{\tau}{; \boldsymbol{\theta}}  \right).
		\label{eq:mapping_stepSize_dnn}
	\end{equation}
	%
	% TODO: add keyword "step-size", "step-size estimation" or "step-size control"
	%
	The parameter vector $\boldsymbol{\theta}$ of the function $ f_{\text{ML}}$ is optimized in a training phase (cf.~Sec.~\ref{sec:costFunc}). %  by minimizing the cost function $\mathcal{J}$ (cf.~Eq.~\eqref{eq:systMisDef_td_av} in Sec.~\ref{sec:costFunc})
	However, a direct estimation of $\boldsymbol{\Lambda}_{\tau}^{\text{DNN}}$ by a \acs{DNN} is complicated by the non-stationarity \commentTHd{and non-whiteness} of the respective signals. Thus, we suggest to exploit the domain knowledge from optimum model-based adaptation control (cf.~Sec.~\ref{sec:impSigStatForAdCon}), by using the following \ac{DFT} bin-wise step-size
	\begin{equation}
		\left[\boldsymbol{\Lambda}_{\tau}^{\text{DNN}} \right]_{mm} = \frac{ \mu_{\text{MAX}} \left[\boldsymbol{M}_{ \tau}^{\mu}\right]_{mm}}{ \left[\hat{\bPsi}_{\tau}^{\text{XX}} + \frac{M}{R} {\hat{\bPsi}{}_{\tau}^{\text{PP}}} \right]_{mm} } \commentTHf{,}
		\label{eq:dnn_step_size}
	\end{equation}
	with
	\begin{align}
		\hat{\bPsi}{}_{\tau}^{\text{XX}} &= \lambda_{\text{X}} ~\hat{\bPsi}{}_{\tau-1}^{\text{XX}} + (1-\lambda_{\text{X}}) ~ {\bx}_{\tau}  {\bx}_{\tau}^{\text{H}}
		\label{eq:inputSigPSDESt} \\
		{\hat{\bPsi}{}_{\tau}^{\text{PP}}} &= {\lambda_{\text{P}}~\hat{\bPsi}{}_{\tau-1}^{\text{PP}} + (1-\lambda_{\text{P}})~ {\hat{\bp}}_\tau  {\hat{\bp}}_\tau  ^{\herm}}
		\label{eq:obs_noise_est_S} \\
		{\hat{\bp}_\tau} &={\boldsymbol{M}}_{\tau}^{e} {\be}_\tau \commentTHf{,}
		\label{eq:noiseEstSig}
	\end{align}
	{where} the diagonal masking matrices ${\boldsymbol{M}}_{\tau}^{\mu}$ and ${\boldsymbol{M}}_{\tau}^e$ {are} {inferred} by the \acs{DNN} {and where $\lambda_{\text{X}}$ and $\lambda_{\text{P}}$ are time constants for recursive averaging.}
	By embedding the signal power normalization into the structure of the learning-based step-size mapping \eqref{eq:mapping_stepSize_dnn}, the \acs{DNN} needs to model a significantly reduced dynamic range in contrast to directly estimating $\boldsymbol{\Lambda}_{\tau}^{\text{DNN}}$. 
	The step-size \eqref{eq:dnn_step_size} is motivated by adding an error power-dependent normalization term $\hat{\bPsi}{}_{\tau}^{\text{PP}}$, similar to the noise \ac{PSD} matrix ${\bPsi}_{\tau}^{\text{NN}}$ in the \ac{KF} update \eqref{eq:stepSizeOptKf}, to the \acs{FDAF} step-size \eqref{eq:fdaf_stepSize} and estimating the decisive hyperparameter $\mu_{\text{FDAF}}$ \commentTHg{frequency-selectively by a} \acs{DNN}. {To account for the different causes of large error powers, i.e., observation noise $\bn_{\tau}$ or system \commentTHe{mismatch} $\Delta \bw_{\tau}$, and their antipodal effect on the adaptation rate, a \acs{DNN}-estimated mask $\boldsymbol{M}_\tau^{e}$ is applied to the error signal $\be_{\tau}$ before computing $\hat{\bPsi}{}_{\tau}^{\text{PP}}$ (cf.~Eqs.~\eqref{eq:obs_noise_est_S} and \eqref{eq:noiseEstSig})}. Note that despite the structural similarity of the step-sizes \eqref{eq:stepSizeOptKf} and \eqref{eq:dnn_step_size}, ${\hat{\bPsi}{}_{\tau}^{\text{PP}}}$ is not necessarily an estimate of the noise signal \ac{PSD} matrix ${{\bPsi}_{\tau}^{\text{NN}}}$. Its actual {meaning} depends on the cost function that is used to train the \acs{DNN}. %(cf.~Sec.~\ref{sec:costFunc}).
	%
	% TODO: add keyword "step-size", "step-size estimation" or "step-size control"
	%
	% TODO: das für bezug zu Rausch PSD am ende
	%
	% TODO: was vereinfacht sich bei den oben genannten schwierigkeiten
	% TODO: PSI_nn nicht unbbeding rausch PSD
	% TODO: warum ist es gut die struktur von (12) aus den modell-basierten Verfahren zu nehmen?
	%
	
	We suggest the \acs{DNN} architecture shown in Fig.~\ref{fig:nn_architecture} to map the {observed} feature vector ${\bu}_{\text{feat},\tau}$ (cf.~Eq.~\eqref{eq:inFeatDNN} \commentTHg{below}) to the diagonal masking matrices ${\boldsymbol{M}}_{\tau}^{\mu}$ and ${\boldsymbol{M}}_{\tau}^e$ required in Eqs.~\eqref{eq:dnn_step_size} and \eqref{eq:noiseEstSig}. The architecture is motivated by the creation of a condensed feature representation after the \ac{GRU} layer which includes all important effects influencing the adaptation control, e.g., noise activity and filter convergence state. For the input feature vector {to the \acs{DNN}} we use the normalized logarithmic power spectrum of the {input} signal ${\bx}_\tau$ and the prior error signal ${\be}_\tau$ to obtain
		\begin{algorithm}[b]
		\caption{\commentTHd{Proposed \ac{DNN-FDAF} update for one signal block.}} %  block of microphone samples $\by_{\tau}$
		%\caption{Algorithmic description of the filter update of the proposed \ac{DNN-FDAF} algorithm.}
		%\caption{Proposed \ac{DNN-FDAF} algorithm for online system identification using $\commentTHb{T_{\text{test}}}$ sequential testing signal blocks.}
		\label{alg:prop_alg_descr}
		\begin{algorithmic}
		%	\For{$\tau=1,\dots,{T_{\text{test}}}$}
			\State Compute prior error block ${\be}_\tau$ by Eq.~\eqref{eq:errComp}
			\State Compute feature vector ${\boldsymbol{u}}_{\text{feat},\tau}$ for the \acs{DNN}  by Eq.~\eqref{eq:inFeatDNN}
			\State Infer masking matrices ${\boldsymbol{M}}_{\tau}^{\mu}$ and ${\boldsymbol{M}}_{\tau}^e$ (cf.~Fig.~\ref{fig:nn_architecture}) 
			\State Compute step-size matrix $\boldsymbol{\Lambda}_{\tau}^{\text{DNN}}$ {by Eqs.~\eqref{eq:dnn_step_size} - \eqref{eq:noiseEstSig}}% \eqref{eq:inputSigPSDESt}, \eqref{eq:dnn_step_size}, \eqref{eq:obs_noise_est_S} 
			\State Update {\ac{FR}} estimate $\hat{\boldsymbol{w}}_{\tau }$ by Eq.~\eqref{eq:filtUpd}
		%	\EndFor
		\end{algorithmic}
	\end{algorithm}
	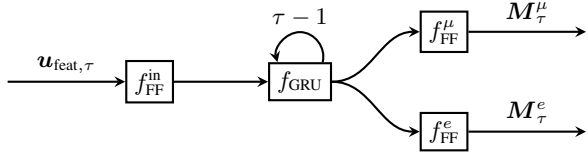
\begin{figure}[t]
		% \vspace*{-.17cm}
		\centering
			
	\begin{tikzpicture}[node distance=2.0cm, >=stealth]

	\node (featSig) at (0,0) {};
	\node [rectangle, draw, thick, right of=featSig] (ffIn) { \commentTHb{$f_{\text{FF}}^{\text{in}}$} };
	\node [rectangle, draw, thick, right of=ffIn] (ffGru) {$f_{\text{GRU}}$};
	\node [rectangle, draw, thick, above right of=ffGru, yshift=-.75cm, xshift=.5cm] (ffOutMu) {\commentTHb{$f_{\text{FF}}^{\mu}$}};
	\node [rectangle, draw, thick, below right of=ffGru, yshift=.75cm, xshift=.5cm] (ffOutE) {\commentTHb{$f_{\text{FF}}^{e}$}};
	\node [right of=ffOutMu] (maskMu) {};
	\node [right of=ffOutE] (maskE) {};
	
	\draw [thick, ->] (featSig) -- (ffIn.west) node [midway, above] {${\bu}_{\text{feat},\tau}$};
	\draw [thick, ->] (ffIn.east) -- (ffGru.west);
	\draw [thick, ->] (ffGru.east) to [out=0,in=180] (ffOutMu.west);
	\draw [thick, ->] (ffGru.east) to [out=0,in=180] (ffOutE.west);
	\draw [thick, ->] (ffOutMu.east) -- (maskMu) node [midway, above] {$\boldsymbol{M}_{ \tau}^{{\mu}}$};
	\draw [thick, ->] (ffOutE.east) -- (maskE) node [midway, above] {$\boldsymbol{M}_{ \tau}^{{e}}$};

	\draw [thick, ->] ($(ffGru.north)+(.3,0)$) to [out=75,in=0] ($(ffGru.north)+(.0,.4)$) to [out=180,in=105] ($(ffGru.north)+(-.3,0)$) ;
	% \draw ($(ffGru.north)+(.3,0)$) arc (0:170:.2);
	\node (delay) at ($(ffGru.north)+(.0,.6)$)  {$\tau-1$};
	
	\end{tikzpicture}
		\caption{Proposed \acs{DNN} architecture which maps the feature vector ${\bu}_{\text{feat},\tau}$ to the diagonal masking matrices $\boldsymbol{M}_{ \tau}^{{\mu}}$ and $\boldsymbol{M}_{ \tau}^{{e}}$.}% used in the step-size matrix computation \eqref{eq:dnn_step_size} - \eqref{eq:noiseEstSig}.}	
		\label{fig:nn_architecture}
		% \vspace*{-.4cm}
	\end{figure}
	\begin{equation}
		%[\tilde{\boldsymbol{u}}_{\text{feat},\tau}]_m = \frac{\text{max}(\log (| [\tilde{\boldsymbol{u}}_{,\tau}]_m|^{2}), \epsilon) - [\boldsymbol{\mu}]_{m}}{ [\boldsymbol{\sigma}]_m},
		[{\boldsymbol{u}}_{\text{feat},\tau}]_m = \frac{{\log {\text{max}(| [{\boldsymbol{u}}_{\text{sig},\tau}]_m|^{2}, \epsilon)}} - [\boldsymbol{\nu}]_{m}}{ [\boldsymbol{\sigma}]_m}
		\label{eq:inFeatDNN}
	\end{equation}
	{which is computed from the complex signal vector}
	\begin{align}
%		{\bu}_{\text{sig},\tau} = \begin{bmatrix} \boldsymbol{Q}_{4}~
%			{{\be}}_{\tau} \\
%			\boldsymbol{Q}_{4}~{\boldsymbol{x}}_{\tau}
%		\end{bmatrix} \in \mathbb{C}^{M+2}.
%		\label{eq:feature_e_x}
		\commentTHd{{\bu}_{\text{sig},\tau} = \begin{bmatrix} \left(\boldsymbol{Q}_{4}~
{{\be}}_{\tau} \right)^{\text{T}} &
\left(\boldsymbol{Q}_{4}~{\boldsymbol{x}}_{\tau}  \right)^{\text{T}}
\end{bmatrix}^{\text{T}} \in \mathbb{C}^{M+2}.}
\label{eq:feature_e_x}
	\end{align}
	%
	% The mean and element-wise standard deviation vectors of the feature vector ${\boldsymbol{u}}_{\text{feat},\tau}$ are denoted by $\boldsymbol{\nu}$ and $\boldsymbol{\sigma}$, respectively, and can be estimated from the training data.
	%\commentTHd{with $\boldsymbol{\nu}$ and $\boldsymbol{\sigma}$ denoting the estimated mean and element-wise standard deviation vector of the logarithmic power spectra.}
	%
	% \commentTHg{The matrix \makebox{$\boldsymbol{Q}_{4} = \begin{pmatrix}\boldsymbol{I}_{\frac{M}{2}+1}&\boldsymbol{0}_{\frac{M}{2}+1 \times \frac{M}{2}-1} \end{pmatrix}$} selects the non-redundant part of the conjugate symmetric signals in \eqref{eq:feature_e_x}, $\epsilon>0$ ensures a positive argument of the logarithm and $\boldsymbol{\nu}$ and $\boldsymbol{\sigma}$ denote the mean and element-wise standard deviation vectors of the logarithmic power spectra.}
	%
	\commentTHd{The mean and element-wise standard deviation vectors of the logarithmic power spectra are denoted by $\boldsymbol{\nu}$ and $\boldsymbol{\sigma}$, respectively.}
	 % The mean and element-wise standard deviation vectors of the feature vector ${\boldsymbol{u}}_{\text{feat},\tau}$ are denoted by $\boldsymbol{\nu}$ and $\boldsymbol{\sigma}$, respectively, and can be estimated from the training data.
%
%	 \commentTHd{As the prior error $\be_{\tau}$ is not known in advance its respective standardization }
%	 
%	%  \commentTHd{The elements of the feature vector ${\boldsymbol{u}}_{\text{feat},\tau}$ are standardized by using the mean and element-wise standard deviation vector
%	 
% }
	  \commentTHd{Furthermore,} the matrix \makebox{$\boldsymbol{Q}_{4} = \begin{pmatrix}\boldsymbol{I}_{\frac{M}{2}+1}&\boldsymbol{0}_{\frac{M}{2}+1 \times \frac{M}{2}-1} \end{pmatrix}$} selects the non-redundant part of the conjugate symmetric signals in \eqref{eq:feature_e_x} and $\epsilon>0$ ensures a {positive} argument of the logarithm. 
	  The feature vector ${\bu}_{\text{feat},\tau}$ is mapped by a feedforward layer with tanh activation {$f_{\text{FF}}^{\text{in}}$} to a lower dimension $P$. Subsequently, two stacked \ac{GRU} layers $f_{\text{GRU}}$ extract temporal dependencies of the compressed feature vectors. Finally, the \ac{GRU} states are mapped by two different feedforward networks {$f_{\text{FF}}^{\mu}$ and $f_{\text{FF}}^{e}$} with sigmoid activations to the {diagonal entries of the} masking matrices ${\boldsymbol{M}}_{\tau}^{\mu}$ and ${\boldsymbol{M}}_{\tau}^e$. The sigmoid activations at the output layers ensure that all elements of the masking matrices lie in the range $[0,~1]$. This contributes to the robustness of the approach by limiting the numerator \commentTHf{of \eqref{eq:dnn_step_size}} to $\mu_{\text{MAX}}$ and the norm of $\hat{{\bp}}_\tau$ (cf.~Eq.~\eqref{eq:noiseEstSig}) to {$||{\be}_\tau ||_2$}. Furthermore, note that due to the conjugate symmetry of the \ac{DFT}-domain representation it suffices to compute the nonredundant part of the masking matrices. An algorithmic description of the proposed \ac{DNN-FDAF} \commentTHd{update} is given in Alg.~\ref{alg:prop_alg_descr}. % ${\boldsymbol{M}}_{\tau}^{\mu}$ and ${\boldsymbol{M}}_{\tau}^e$.} 
	
	\subsection{Cost Function Design for Neural Network Training}
	\label{sec:costFunc}
\begin{figure}[t]
	% \vspace*{-.2cm}
	\centering
		
	\begin{tikzpicture}[node distance=1.49cm, >=stealth]

	\def\ySpaceAA{.4}
	\def\ySpaceBB{-.2}
	\def\innerSepAA{.05}
	\def\innerSepB{.05}
	\def\innerSepC{.1}
	
	\node [circle, draw,inner sep=\innerSepB cm, thick] (hatW0) at (0,0) {$\hat{\bw}_{j,0}$};

	\node [circle, draw, inner sep=\innerSepB cm, right of=hatW0, thick] (hatW1)  {$\hat{\bw}_{j,1}$};
 	\node [circle, draw, inner sep=\innerSepB cm, right of=hatW1, thick] (hatW2) {$\hat{\bw}_{j,2}$};
	\node [right of=hatW2] (hatW3) {\large$\dots$};
	\node [circle, draw, inner sep=\innerSepB cm, right of=hatW3, thick] (hatW4) {$\hat{\bw}_{j,T}$};

	\node [circle, draw,inner sep=\innerSepAA cm, right of=hatW0, above of=hatW0, yshift=\ySpaceBB cm, thick] (lambda1) {$\boldsymbol{\Lambda}_{j,1}^{\text{DNN}} $};
	\node [circle, draw,inner sep=\innerSepAA cm, right of=hatW1, above of=hatW1, yshift=\ySpaceBB cm, thick] (lambda2) {$\boldsymbol{\Lambda}_{j,2}^{\text{DNN}} $};
	\node [right of=lambda2,yshift=0 cm] (lambda3) {\large$\dots$};
	\node [circle, draw,inner sep=\innerSepAA cm, right of=lambda3, yshift=0 cm, thick] (lambda4) {$\boldsymbol{\Lambda}_{j,T}^{\text{DNN}} $};
	
	\node [rectangle, draw,inner sep=\innerSepC cm, right of=hatW0, below of=hatW0, yshift=\ySpaceAA cm, thick] (ups1) {\commentTHa{${{\Upsilon}}_{j,1}$}};
	\node [rectangle, draw,inner sep=\innerSepC cm, right of=hatW1, below of=hatW1, yshift= \ySpaceAA cm, thick] (ups2) {\commentTHa{${{\Upsilon}}_{j,2}$}};
	\node [right of=ups2, yshift=.0cm] (ups3) {\large$\dots$};
	\node [rectangle, draw,inner sep=\innerSepC cm, right of=ups3, yshift=.0cm, thick] (ups4) {\commentTHa{${{\Upsilon}}_{j,T}$}};
	
	\node [circle, draw,inner sep=\innerSepAA cm, below of=ups1, yshift=\ySpaceAA cm, thick] (W1) {${\bw}_{j,1}$};
	\node [circle, draw,inner sep=\innerSepAA cm,  below of=ups2, yshift=\ySpaceAA cm, thick] (W2) {${\bw}_{j,2}$};
	\node [right of=W2, thick] (W3) {\large$\dots$};
	\node [circle, draw,inner sep=\innerSepAA cm,  below of=ups4, yshift=\ySpaceAA cm, thick] (W4) {${\bw}_{j,T}$};
	
	\node [circle, dashed, draw,inner sep=0.10cm,  above of=lambda1, xshift=2.25cm, yshift=-.4cm, thick] (theta) {$\boldsymbol{\theta}$};
	
	\draw [thick, ->] (hatW0) -- (hatW1);
	\draw [thick, ->] (hatW1) -- (hatW2);
	\draw [thick, ->] (hatW2) -- (hatW3);
	\draw [thick, ->] (hatW3) -- (hatW4);
	\draw [thick, ->] (lambda1) -- (hatW1);
	\draw [thick, ->] (lambda2) -- (hatW2);
	\draw [thick, ->] (lambda4) -- (hatW4);
	
	\draw [thick, ->] (W1) -- (ups1);
	\draw [thick, ->] (W2) -- (ups2);
	% \draw [thick, ->] (W3) -- (ups3);
	\draw [thick, ->] (W4) -- (ups4);
	
	\draw [thick, ->] (hatW1) -- (ups1);
	\draw [thick, ->] (hatW2) -- (ups2);
	%\draw [thick, ->] (hatW3) -- (ups3);
	\draw [thick, ->] (hatW4) -- (ups4);
	
	\draw [thick, ->] (hatW0) -- (lambda1);
	\draw [thick, ->] (hatW1) -- (lambda2);
	\draw [thick, ->] (hatW2) -- (lambda3);
	\draw [thick, ->] (hatW3) -- (lambda4);
	
	\draw [thick, dashed, ->] (theta) -- (lambda1);
	\draw [thick, dashed, ->] (theta) -- (lambda2);
	\draw [thick, dashed, ->] (theta) -- (lambda3);
	\draw [thick, dashed, ->] (theta) -- (lambda4);
	
	\end{tikzpicture}
	\caption{Visualizing the relationship between the cost function terms {${{\Upsilon}}_{j,\tau}$} and the \commentTHf{\acs{DNN} parameter vector} $\boldsymbol{\theta}$.}% {Note that for an improved clarity we omitted the training sequence index $j$.}	
	\label{fig:back_prop_gradients}
\end{figure}
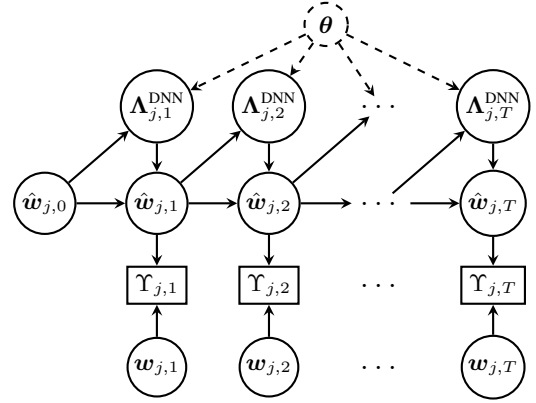

	The system identification performance of an adaptive filter is often quantified by \commentTHd{the \commentTHd{\ac{NESD}}} \cite{enzner_acoustic_2014}
	%
%	\vspace*{-.2cm}	
	\begin{equation}
		% \Upsilon_{\tau} = 10 \log_{10} \frac{||{\underline{\bw}}_{\tau} - \hat{\underline{\bw}}_{\tau}||_{2}^2}{||\underline{\bw}_{\tau}||_{2}^2}.
		\Upsilon_{\tau} = \frac{||{\underline{\bw}}_{\tau} - \hat{\underline{\bw}}_{\tau}||_{2}^2}{||\underline{\bw}_{\tau}||_{2}^2}.
		\label{eq:systMisDef_td}
	%	\vspace*{-.2cm}	
	\end{equation}
	%
	%
	%\commentTHd{shortly termed system distance in the following.}
	% TODO: check if this should be deleted
	% Note that $\Upsilon_{\tau}$ is usually called distance although it does not fulfil the properties of a metric on the vector space $\mathbb{R}^L$. 
	\commentTHf{Due} to the complex interaction of the \acs{DNN} outputs, i.e., the masking matrices ${\boldsymbol{M}}_{\tau}^{\mu}$ and ${\boldsymbol{M}}_{\tau}^e$ (cf.~Fig.~\ref{fig:nn_architecture}), via a sequence of filter updates (cf. Eqs.~\eqref{eq:filtUpd} and \eqref{eq:dnn_step_size} - \eqref{eq:noiseEstSig}) on the \commentTHd{\ac{NESD}} $\Upsilon_{\tau}$, a hand-crafted design of optimum target masking matrices is problematic. Thus, we suggest an end-to-end approach by directly optimizing the \acs{DNN} parameter vector $\boldsymbol{\theta}$ w.r.t. to the \commentTHd{average logarithmic \commentTHd{\ac{NESD}}}
	%
% 	\vspace*{-.2cm}	
	\begin{equation}
		% {\bar{\Upsilon}(\boldsymbol{\theta}) = \frac{1}{J T } \sum_{j=1}^J\sum_{\tau=1}^{T} \Upsilon_{j,\tau}} {,}
		% {\mathcal{J}(\boldsymbol{\theta})} = \frac{1}{J T } \sum_{j=1}^J\sum_{\tau=1}^{T} \Upsilon_{j,\tau} {,}
		{\mathcal{J}(\boldsymbol{\theta})} = \frac{1}{J T } \sum_{j=1}^J\sum_{\tau=1}^{T} 10 \log_{10} \left(  \Upsilon_{j,\tau} \right) {,}
		\label{eq:systMisDef_td_av}
	%	\vspace*{-.2cm}	
	\end{equation}
	with $J$ and $T$ being the number of training sequences and signal blocks, respectively, and $\Upsilon_{j,\tau}$ denoting the \commentTHd{\ac{NESD}} \eqref{eq:systMisDef_td} at block $\tau$ in training sequence $j$. The cost function \eqref{eq:systMisDef_td_av} quantifies the direct effect of different masking matrices on the average system identification performance of the adaptive filter and renders the design of desired target signal statistics and \commentTHg{choice of critical} hyperparameters unnecessary. The end-to-end training of the \acs{DNN} requires to backpropagate the average \commentTHd{\ac{NESD}} \eqref{eq:systMisDef_td_av} through the adaptive filter updates \eqref{eq:filtUpd} to the \acs{DNN} parameter vector $\boldsymbol{\theta}$. This {complex} relation between the cost function terms $\Upsilon_{j,\tau}$, the {\ac{FR}} estimates $\hat{\bw}_{j,\tau}$, the step-size matrices {$\boldsymbol{\Lambda}_{j,\tau}^{\text{DNN}}$} and the \acs{DNN} parameters $\boldsymbol{\theta}$ is shown in Fig.~\ref{fig:back_prop_gradients}. Finally, note that due to the \commentTHj{lack of explicit dependency} of the \commentTHd{\ac{NESD}} \eqref{eq:systMisDef_td} on the signal characteristics, the cost function \eqref{eq:systMisDef_td_av} is well-suited to quantify the system identification performance for non-stationary input signals as typically encountered in acoustic applications.
	
	\section{Experiments}
	\label{sec:experiments}
	In this section, we evaluate the proposed algorithm for a large variety of challenging acoustic system identification scenarios which are motivated by an \ac{AEC} application \commentTHj{with continuous double-talk}. The scenarios are characterized by abrupt {changes of \acp{AIR}} and non-stationary and non-white input $\underline{\bx}_{\tau}$ and noise signals $\underline{\bn}_{\tau}$.
	The noise-free observation {component} $\underline{\bd}_{\tau}$ of each scenario is simulated by randomly drawing an {input} signal $\underline{\bx}_\tau$ from a subset of \commentTHj{the \textit{LibriSpeech} database} \cite{7178964}, including $143$ speakers, and convolving it with a randomly-selected true \ac{AIR} $\underline{\tilde{\bw}}_\tau \in \mathbb{R}^{K}$ from the databases \cite{jeub_binaural_2009,Wen06evaluationof, mird}, comprising $201$ different \acp{AIR} \commentTHj{with reverberation times ${T}_{60}$ $\in [ 120 \text{ms},~780 \text{ms}]$}.
	%\commentTHj{with reverberation times $\text{T}_{60}$ ranging from $120$ ms to $780$ ms}.
	%
	Note that, as the length $K$ of the true \acp{AIR} $\underline{\tilde{\bw}}_\tau$ is much larger than the modeled filter length $L$ in all considered scenarios, we can only estimate the first $L$ taps of $\underline{\tilde{\bw}}_\tau$, i.e., {we choose} $\underline{{\bw}}_\tau = \boldsymbol{Q}_5^{\text{T}}  \underline{\tilde{\bw}}_\tau$ with \makebox{$\boldsymbol{Q}_5^{\text{T}} = \begin{pmatrix}\boldsymbol{I}_{L}&\boldsymbol{0}_{L \times K-L} \end{pmatrix} $} {in Eq.~\eqref{eq:systMisDef_td}}.
	Subsequently, the microphone observation $\underline{\by}_\tau$ is computed by adding a noise signal $\underline{\bn}_\tau$. Each noise signal consists of a superposition of a randomly selected speaker from a \commentTHf{disjoint} subset of \cite{7178964}, including $145$ speakers, and a stationary white Gaussian signal.
	Both noise components are scaled to yield a random \acs{SNR}, i.e., power of the noise-free component $\underline{\bd}_{\tau}$ w.r.t. the noise component, between $-10$ dB and $10$ dB for the \commentTHj{interfering} speaker, and $25$ dB and $35$ dB for the Gaussian component.
	The abrupt system change is modeled by using different \acp{AIR}, {input} and noise signals for creating the observations $\underline{\by}_\tau$ before and after a specific switching time. We sampled the switching time randomly in the range $\left[7.2\text{s},~8.8\text{s}\right]$ to {preclude} overfitting of the \acs{DNN} to a deterministic point in time.
	%
	%\commentTHj{The switching time was sampled randomly in the range $\left[7.2\text{s},~8.8\text{s}\right]$ to preclude overfitting of the \acs{DNN} to a deterministic point in time.}
	%
 	%
	 \begin{table}[b]
	 	\vspace*{.1cm}
	 	\caption{Parameter settings for the considered algorithms.}\vspace{-0mm}% components of the proposed \ac{AEC} algorithm. For the best performance values bold font is used.}\vspace{-0mm}
	 	% Performance evaluation of the various algorithmic parts for the proposed algorithm. 
	 	\vspace*{-.35cm} % \vspace*{-.35cm}
	 	\setlength{\tabcolsep}{10.0pt}
	 	\begin{center}
	 		\begin{tabular}{l c c c}
	 			\toprule
	 			Algorithm																			& $\lambda_{\text{X}}$ & $ {\lambda_{\text{P}}} $ & $\mu_{\text{MAX}}$  \\ \midrule
	 			% KF 																					& -- 	& $0.5$ & -- \\ 
	 			{EA}-FDAF 																			& $0.5$ & $0.5$ & $0.75$ \\ 
	 			DNN-FDAF ({${\boldsymbol{M}}_{\tau}^{e}=\boldsymbol{0}_{M \times M}$}) 	& $0.5$ & $0.0$ & $1.0$  \\ 
	 			DNN-FDAF (${\boldsymbol{M}}_{\tau}^{\mu}=\boldsymbol{I}_M$) 						& $0.5$ & $0.0$ & $0.5$  \\ 
	 			DNN-FDAF 																			& $0.5$ & $0.0$ & $1.0$ \\
	 			\bottomrule 
	 		\end{tabular} 
	 		\vspace{-.35cm}
	 		%\vspace{1.05cm}
	 	\end{center}
	 	\label{tab:tabParSet}
	 	% \vspace{-.3cm}
	 	% DNN-FDAF (M^e=0) ==   'fdaf_nn_GRU_(256,1)_recAvGrad_False_stepSize_True(0.5,1.0,False,0.5)_recAvHf_False_maskEfNum_False_denMic_False_systMis_tempWeight_None_Xf-Ef_mat'
	 	% DNN-FDAF (M^mu=I) ==  'fdaf_nn_GRU_(256,1)_recAvGrad_False_stepSize_False(0.5,None,False,0.5)_recAvHf_False_maskEfNum_True_denMic_False_systMis_tempWeight_None_Xf-Ef_mat'
	 	% DNN-FDAF == 			'fdaf_nn_GRU_(256,1)_recAvGrad_False_stepSize_True(0.5,1.0,False,0.5)_recAvHf_False_maskEfNum_True_denMic_False_systMis_tempWeight_None_Xf-Ef_mat'   
	 \end{table}

	For all considered algorithms the sampling frequency $f_s$ is $16$ kHz and the modeled filter length and frame shift are set to $L=2048$ and $R=1024$, respectively. 
	As baseline algorithms we consider the \ac{KF} approach \cite{enzner_frequency-domain_2006} and an error-aware version of the \acs{FDAF} step-size \eqref{eq:fdaf_stepSize} which is termed EA-FDAF, i.e., setting ${\boldsymbol{M}}_{\tau}^{\mu}={\boldsymbol{M}}_{\tau}^e=\boldsymbol{I}_M$ in \commentTHj{Eqs.}~\eqref{eq:dnn_step_size} and \eqref{eq:noiseEstSig}. Note that for the \ac{KF} update the noise \ac{PSD} matrix is computed by recursively averaging the prior error ${\be}_\tau$ with an averaging factor of $0.5$ \cite{malik_online_2010, franzen_improved_2019}.
	% , i.e., using \eqref{eq:obs_noise_est_S} with {${\hat{\bn}}_\tau = {\be}_\tau$} . 
	%In addition, we consider two choices of the state transition parameter $A$, i.e., $A=0.99$ and $A=0.999$, as it highly affects the steady-state and reconvergence performance \cite{yang_frequency-domain_2017}. 
	\commentTHj{We consider two choices of the state transition parameter $A$ controlling the time constant of \ac{FIR} filter changes. While a higher $A$ gives better steady-state performance, a lower $A$ favors fast reconvergence~\cite{yang_frequency-domain_2017}.}
	For the proposed {\ac{DNN-FDAF} algorithm we consider two additional variants which are termed \makebox{\ac{DNN-FDAF} ({${\boldsymbol{M}}_{\tau}^{e}=\boldsymbol{0}_{M \times M}$})} and \makebox{\ac{DNN-FDAF} (${\boldsymbol{M}}_{\tau}^{\mu}=\boldsymbol{I}_M$)}}. Here, the respective quantity in brackets is {fixed}, i.e., {it} is not estimated by the \acs{DNN}. The parameter settings of the considered algorithms are summarized in Tab.~\ref{tab:tabParSet}. Note that the maximum numerator step-size $\mu_{\text{MAX}}$ is chosen smaller for the \ac{DNN-FDAF} (${\boldsymbol{M}}_{\tau}^{\mu}=\boldsymbol{I}_{M}$) algorithm {to compensate for} the deterministic numerator of the step-size \eqref{eq:dnn_step_size}. 
	The considered \acs{DNN} architecture (cf.~Fig.~\ref{fig:nn_architecture}) has approximately $2.4$ million parameters with \commentTHd{the number of hidden GRU states at each layer being $P=256$} and $\epsilon=10^{-12}$. It was trained on $4.4$ h of training data using the ADAM optimizer \cite{kingma2014adam} with a learning rate of $10^{-3}$. The normalization variables $\boldsymbol{\nu}$ and $\boldsymbol{\sigma}$ in the feature computation \eqref{eq:inFeatDNN} are estimated from the training data with the prior error statistics being approximated by the respective microphone signal statistics.
	% The considered \acs{DNN} architecture (cf.~Fig.~\ref{fig:nn_architecture}) has approximately $2.4$ million parameters with the output dimension after the first dense layer being $P=256$ and $\epsilon=10^{-12}$. It was trained on $4.4$ h of training data using the ADAM optimizer \cite{kingma2014adam} with a learning rate of $10^{-3}$. 
	
	% As performance measures we consider the average logarithmic {\ac{ERLE}} \cite{enzner_acoustic_2014}
	%
%	\begin{equation}
%		\bar{\mathcal{E}}_{\tau} = \frac{1}{I} \sum_{i=1}^{I} 10 \log_{10} \frac{\mathbb{E}\left[||\underline{\boldsymbol{d}}_{i,\tau}||^2_2\right] }{\mathbb{E}\left[||\underline{\boldsymbol{d}}_{i,\tau}-\widehat{\boldsymbol{\underline{d}}}_{i,\tau}||^2_2\right]}
%		\label{eq:erleDef}
%	\end{equation}
%	%
%	with the expectation {operator $\mathbb{E}$} being approximated by recursive averaging and the \commentTHd{average \commentTHd{logarithmic} zero-padded \commentTHd{\ac{NESD}}} \cite{enzner_acoustic_2014}
%	%
	%
	We consider the average logarithmic zero-padded \ac{NESD} \cite{enzner_acoustic_2014}
	\begin{equation}
	\vspace*{.15cm}
	{\bar{\Upsilon}_{\text{ZP},\tau}} = \frac{1}{I} \sum_{i=1}^{I}  10 \log_{10} \frac{||  {\tilde{\underline{\bw}}}_{i,\tau} -\boldsymbol{Q}_5 \hat{\underline{\bw}}_{i,\tau}||_{2}^2}{||\tilde{\underline{\bw}}_{i,\tau}||_{2}^2}
	\label{eq:systMisDef_td_long}
	\end{equation}	
	and the average logarithmic {\acl{ERLE}} \cite{enzner_acoustic_2014}
	\begin{equation}
		\bar{\mathcal{E}}_{\tau} = \frac{1}{I} \sum_{i=1}^{I} 10 \log_{10} \frac{\mathbb{E}\left[||\underline{\boldsymbol{d}}_{i,\tau}||^2_2\right] }{\mathbb{E}\left[||\underline{\boldsymbol{d}}_{i,\tau}-\widehat{\boldsymbol{\underline{d}}}_{i,\tau}||^2_2\right]}
		\label{eq:erleDef}
	\end{equation}
	as performance measures. %with the expectation {operator $\mathbb{E}$} in \eqref{eq:erleDef} being approximated by recursive averaging \commentTHj{to provide a time-dependent metric}.
	\commentTHj{Note that a zero-padded version of the estimate $\hat{\underline{\bw}}_{i,\tau}$ is used in \eqref{eq:systMisDef_td_long} to account for the undermodeling of the \ac{FIR} filter model \cite{enzner_acoustic_2014} and that we approximate the expectation operator in \eqref{eq:erleDef} by recursive averaging to provide a time-dependent metric.}
	% and $I=100$ ({corresponding to} $27$ min) defining the number of experiments with varying speakers, \acp{AIR} and transition times.}
	%
		\begin{figure}[t!]
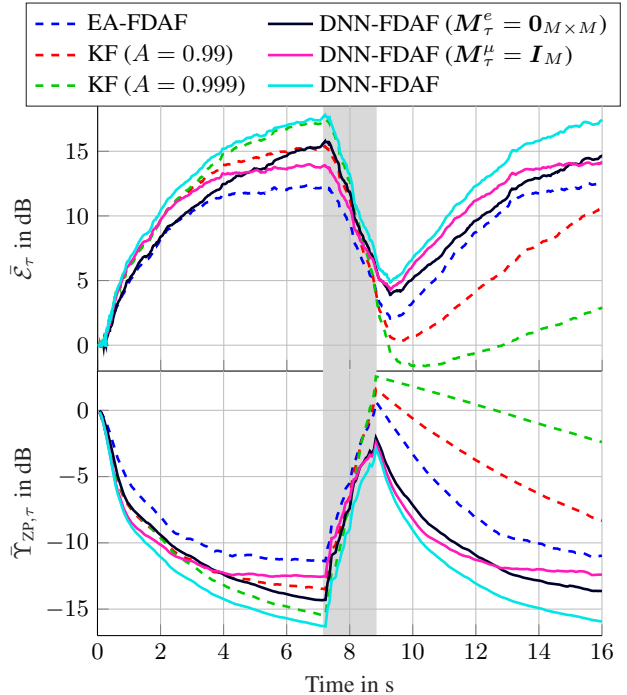

		\centering
		
		\hspace*{.47cm}
		
		\begin{subfigure}[t]{1\columnwidth}
			\centering
			\newlength\fwidth
			\setlength\fwidth{1\columnwidth}
\definecolor{mycolor1}{rgb}{0.00000,0.00000,0.17241}%
\definecolor{mycolor2}{rgb}{1.00000,0.10345,0.72414}%
\definecolor{mycolor3}{rgb}{0.0,0.9,0.9}%
\begin{tikzpicture}

\begin{axis}[%
width=0.0\fwidth,
height=0.0\fwidth,
at={(0\fwidth,0\fwidth)},
scale only axis,
xmin=0.0000,
xmax=16.0000,
xlabel style={font=\color{white!15!black}},
ymin=-16.0000,
ymax=4.0000,
axis background/.style={fill=white},
axis x line*=bottom,
axis y line*=left,
xmajorgrids,
ymajorgrids,
		 transpose legend,
legend style={legend cell align=left, align=left, draw=white!15!black},  legend columns=3
]
\addplot [color=blue, line width=1.0pt, dashed]
table[row sep=crcr]{%
	0.0640	-0.0610\\
};
\addlegendentry{\commentTHb{EA}-FDAF} % fdaf 0.75

\addplot [color=red, line width=1.0pt, dashed]
table[row sep=crcr]{%
	0.0640	-0.1706\\
};
\addlegendentry{KF ($A=0.99$)} % kf 0.99 10

\addplot [color=green!80!black, line width=1.0pt, dashed]
table[row sep=crcr]{%
	0.0640	-0.1706\\
};
\addlegendentry{KF ($A=0.999$)\hspace*{.2cm}} % kf 0.999 10

\addplot [color=mycolor1, line width=1.0pt]
table[row sep=crcr]{%
	0.0640	-0.0166\\
};
% \addlegendentry{fdaf nn GRU (256,1) recAvGrad False stepSize True(0.5,2.0) recAvHf False maskEfNum False denMic False systMis tempWeight None Xf-Ef}
\addlegendentry{DNN-FDAF (\commentTHa{${\boldsymbol{M}}_{\tau}^{e}=\boldsymbol{0}_{M \times M}$})}
% \addlegendentry{DNN-FDAF (${\boldsymbol{M}}_{\tau}^{\mu}$, $\boldsymbol{I}_M$)}

\addplot [color=mycolor2, line width=1.0pt]
table[row sep=crcr]{%
	0.0640	-0.1214\\
};
% \addlegendentry{fdaf nn GRU (256,1) recAvGrad False stepSize False(0.5,None) recAvHf False maskEfNum True denMic False systMis tempWeight None Xf-Ef}
% \addlegendentry{DNN-FDAF ($\boldsymbol{I}_M$, ${\boldsymbol{M}}_{\tau}^e$)}  % ${\boldsymbol{M}}_{\tau}^{\mu}$ and ${\boldsymbol{M}}_{\tau}^e$
\addlegendentry{DNN-FDAF (${\boldsymbol{M}}_{\tau}^{\mu} = \boldsymbol{I}_M$)}  % ${\boldsymbol{M}}_{\tau}^{\mu}$ and ${\boldsymbol{M}}_{\tau}^e$

\addplot [color=mycolor3, line width=1.0pt]
table[row sep=crcr]{%
	0.0640	-0.1497\\
};
% \addlegendentry{fdaf nn GRU (256,1) recAvGrad False stepSize True(0.5,2.0) recAvHf False maskEfNum True denMic False systMis tempWeight None Xf-Ef}
%\addlegendentry{DNN-FDAF (${\boldsymbol{M}}_{\tau}^{\mu}$, ${\boldsymbol{M}}_{\tau}^e$)}
\addlegendentry{DNN-FDAF}  % ${\boldsymbol{M}}_{\tau}^{\mu}$ and ${\boldsymbol{M}}_{\tau}^e$

\end{axis}
\end{tikzpicture}%
\hspace*{-0.2cm}
			
		\end{subfigure}
		
		\vspace*{-.05cm} % \vspace*{.05cm} \vspace*{.35cm}
		
		\begin{subfigure}[t]{\columnwidth}
			\centering
			\setlength\fwidth{.820\columnwidth} % {.85\columnwidth}
			\iftoggle{long}{%
				\hspace*{-.25cm}  \input{erleOn_new.tikz} % \input{images/erleOn_stepSize_2.tikz}
			}{%
				\input{erle.tex}
			}
		\end{subfigure}

		 \vspace*{-0.265cm}% \vspace*{-0.265cm}
		
		%\hspace*{.03cm}
		\begin{subfigure}[t]{\columnwidth}
			\centering
			\setlength\fwidth{.820\columnwidth} % {.85\columnwidth}
			\iftoggle{long}{%
				\input{systMisOn_long_new.tikz} % \input{images/systMisOn_long_stepSize_2.tikz}		
			}{%
				\input{images/systMis.tex}
			}	
		\end{subfigure}
		% \vspace*{-.5cm}
		\vspace*{-.42cm}
		%\caption{\commentTHd{Average performance measures} of the proposed \ac{DNN-FDAF} algorithm, including two algorithmic variants, in comparison to various baselines. The shaded grey area represents the transition period of abrupt system changes.}
		\caption{Performance evaluation of the proposed \ac{DNN-FDAF} algorithm, including \commentTHg{two variants}, \commentTHd{for $I=100$ different scenarios} in comparison to various baselines. \commentTHg{The shaded area represents the period where abrupt system changes occur at random time instants.}
		% The shaded grey area represents the transition period of \commentTHd{randomized} abrupt system changes.
		}
		\label{fig:resResults}
		\vspace*{.02cm}
	\end{figure}
	%
%	\begin{equation}
%	\vspace*{.15cm}
%		{\bar{\Upsilon}_{\text{ZP},\tau}} = \frac{1}{I} \sum_{i=1}^{I}  10 \log_{10} \frac{||  {\tilde{\underline{\bw}}}_{i,\tau} -\boldsymbol{Q}_5 \hat{\underline{\bw}}_{i,\tau}||_{2}^2}{||\tilde{\underline{\bw}}_{i,\tau}||_{2}^2}
%		\label{eq:systMisDef_td_long}
%	\end{equation}
	%
	%with $I=100$ ({corresponding to} $27$ min) being the number of experiments \commentTHd{with varying speakers, \acp{AIR} and transition times}.
	%
	%\commentTHl{Note that we use a zero-padded version of the estimate $\hat{\underline{\bw}}_{i,\tau}$ in \eqref{eq:systMisDef_td_long} to account for the undermodeling of the \commentTHl{\ac{FIR} filter model} \cite{enzner_acoustic_2014}.}
	% 	
	The performance measures ${\bar{\Upsilon}_{\text{ZP},\tau}}$ and $\bar{\mathcal{E}}_{\tau}$ represent arithmetic averages of $I=100$ different experiments ({corresponding to} $27$ min) with varying transition times, speakers and \acp{AIR} which were disjoint from the training data.
%	
%and $I=100$ ({corresponding to} $27$ min) defining the number of experiments with varying speakers, \acp{AIR} and transition times.}
%
%}
%	
%	Note that we use a zero-padded version of the estimate $\hat{\underline{\bw}}_{i,\tau}$ \commentTHd{in \eqref{eq:systMisDef_td_long} to account for} the undermodeling of the \commentTHl{\ac{FIR} filter model} \cite{enzner_acoustic_2014}.
	% Note that we use a zero-padded version of the estimate $\hat{\underline{\bw}}_{i,\tau}$ in Eq.~\eqref{eq:systMisDef_td_long} to take into account the undermodeling of the \ac{FIR} filter model \cite{enzner_acoustic_2014}. 
%	The test data was disjoint from the training data, i.e., \acp{AIR} and speakers \commentTHg{were different.}% , i.e., \commentTHd{different} \acp{AIR} and speakers \commentTHg{were used}.
	
	We conclude from Fig.~\ref{fig:resResults} that the proposed \ac{DNN-FDAF} algorithm significantly outperforms the baselines in terms of convergence rate and reconvergence rate after abrupt \ac{AIR} changes. 
	Furthermore, while either estimating ${\boldsymbol{M}}_{\tau}^{\mu}$ or ${\boldsymbol{M}}_{\tau}^e${, while keeping the other one fixed, results in robust reconvergence at the cost of worse steady-state performance{, their} joint estimation does not need any compromise.} 
	The average \commentTHd{runtime} of the proposed \ac{DNN-FDAF} algorithm for processing one signal block of duration \makebox{$64$ ms} on an \textit{Intel Xeon CPU E3-1275 v6 @ 3.80GHz} is \makebox{$t_{\text{DNN}}=1.5$ ms} which \commentTHg{confirms real-time capability on such platforms.}
	%renders the {method} to be real-time capable \commentTHd{on our machine}.
	% The average inference time of the proposed \ac{DNN-FDAF} algorithm for processing one signal block of duration \makebox{$64$ ms} on an \textit{Intel Xeon CPU E3-1275 v6 @ 3.80GHz} is \makebox{$t_{\text{DNN}}=1.5$ ms} which renders the method to be real-time capable.
	%\vspace*{-.05cm}
	
	\section{Conclusion}
	\label{sec:summaryOutlook}
	%
	% In this paper, we proposed a novel adaptation control for online frequency-domain system identification by using a \acs{DNN} for step-size inference. By optimizing the \acs{DNN} parameters end-to-end w.r.t. the \commentTHd{average \commentTHd{\ac{NESD}}} of the adaptive filter, the proposed \ac{DNN-FDAF} algorithm circumvents the explicit estimation of target signal statistics for model-based step-size estimation.
	% The proposed \ac{DNN-FDAF} algorithm circumvents the explicit design of target signal statistics for model-based step-size estimation by optimizing the \acs{DNN} parameters end-to-end w.r.t. the system {distance} of the adaptive filter. 
	% This renders the method robust against model inaccuracies and high-level {and} non-stationary noise signals and additionally avoids the need for any application-dependent hyperparameter tuning.
	In this paper, we proposed a novel adaptation control for \commentTHd{online system identification} by using a \acs{DNN} for step-size inference. By optimizing the \acs{DNN} parameters end-to-end w.r.t. the \commentTHd{average \commentTHd{\ac{NESD}}} of the adaptive filter, the \commentTHd{proposed algorithm} circumvents the explicit estimation of target signal statistics for model-based step-size estimation.
	This renders the method robust against model inaccuracies and high-level and non-stationary noise signals \commentTHd{and avoids} the need for application-dependent hyperparameter tuning. \commentTHd{Future work may include the coupling of the proposed method with other algorithmic parts of a signal enhancement system, e.g., postfiltering.}
	% TODO: could write an outlook
	
	%
%	\clearpage
%	\newpage
%	
%	
%	\begin{itemize}
%		\item \textcolor{red}{describe freuqency-domain adpative filter for frequency domain step size}
%	\end{itemize}

		\clearpage
	\newpage

\bibliographystyle{IEEEbib}
\bibliography{refs}

\begin{thebibliography}{10}

\bibitem{enzner_acoustic_2014}
G.~Enzner, H.~Buchner, A.~Favrot, and F.~Kuech,
\newblock ``Acoustic {Echo} {Control},''
\newblock in {\em Academic {Press} {Library} in {Signal} {Processing}}, vol.~4,
  pp. 807--877. Elsevier, FL, USA, 2014.

\bibitem{zhang_deep_2019}
H.~Zhang, K.~Tan, and D.~Wang,
\newblock ``Deep {Learning} for {Joint} {Acoustic} {Echo} and {Noise}
  {Cancellation} with {Nonlinear} {Distortions},''
\newblock in {\em Interspeech}, Graz, AT, Sept. 2019, pp. 4255--4259.

\bibitem{westhausen2020acoustic}
N.~L. Westhausen and B.~T. Meyer,
\newblock ``Acoustic echo cancellation with the dual-signal transformation
  {LSTM} network,''
\newblock in {\em Int. Conf. Acoust., Speech, Signal Process.}, Toronto, CA,
  June 2021, pp. 7138--7142.

\bibitem{combAdFiltAndComValDPF}
Mhd.~M. Halimeh, T.~Haubner, A.~Briegleb, A.~Schmidt, and W.~Kellermann,
\newblock ``Combining adaptive filtering and complex-valued deep postfiltering
  for acoustic echo cancellation,''
\newblock in {\em Int. Conf. Acoust., Speech, Signal Process.}, Toronto, CA,
  June 2021, pp. 121--125.

\bibitem{kfNN}
T.~Haubner, Mhd.~M. Halimeh, A.~Brendel, and W.~Kellermann,
\newblock ``A {S}ynergistic {K}alman- and {D}eep {P}ostfiltering {A}pproach to
  {A}coustic {E}cho {C}ancellation,''
\newblock in {\em European Signal Process. Conf.}, Dublin, IE, Aug. 2021.

\bibitem{haykin_2002}
S.~Haykin,
\newblock {\em Adaptive {F}ilter {T}heory},
\newblock Prentice Hall, NJ, USA, 2002.

\bibitem{diniz_adaptive_filtering}
P.~S.~R. Diniz,
\newblock {\em Adaptive Filtering: Algorithms and Practical Implementation},
\newblock Springer, Berlin, Heidelberg, 2007.

\bibitem{haensler2004acoustic}
E.~H{\"a}nsler and G.~Schmidt,
\newblock {\em Acoustic {E}cho and {N}oise {C}ontrol: {A} practical
  {A}pproach},
\newblock Wiley-Interscience, NJ, USA, 2004.

\bibitem{undermodeling}
C.~Paleologou, S.~Ciochină, and J.~Benesty,
\newblock ``Double-talk robust {VSS}-{NLMS} algorithm for under-modeling
  acoustic echo cancellation,''
\newblock in {\em Int. Conf. Acoust., Speech, Signal Process.}, Las Vegas, USA,
  Apr. 2008, pp. 245--248.

\bibitem{gansler_double-talk_1996}
T.~Gansler, M.~Hansson, C.-J. Ivarsson, and G.~Salomonsson,
\newblock ``A double-talk detector based on coherence,''
\newblock {\em IEEE Trans. Commun.}, vol. 44, no. 11, pp. 1421--1427, Nov.
  1996.

\bibitem{Benesty_new_2000}
J.~Benesty, D.~R. Morgan, and J.~H. Cho,
\newblock ``A new class of doubletalk detectors based on cross-correlation,''
\newblock {\em IEEE Trans. Speech Audio Process.}, vol. 8, no. 2, pp. 168--172,
  Mar. 2000.

\bibitem{6112248}
H.~{Huang} and J.~{Lee},
\newblock ``A new variable step-size {NLMS} algorithm and its performance
  analysis,''
\newblock {\em IEEE Trans. Signal Process.}, vol. 60, no. 4, pp. 2055--2060,
  2012.

\bibitem{nitsch2000frequency}
B.~H. Nitsch,
\newblock ``A frequency-selective stepfactor control for an adaptive filter
  algorithm working in the frequency domain,''
\newblock {\em Signal Process.}, vol. 80, no. 9, pp. 1733--1745, 2000.

\bibitem{benesty-vss-lms}
J.~{Benesty}, H.~{Rey}, L.~R. {Vega}, and S.~{Tressens},
\newblock ``A {N}onparametric {VSS} {NLMS} {A}lgorithm,''
\newblock {\em IEEE Signal Process. Lett.}, vol. 13, no. 10, pp. 581--584,
  2006.

\bibitem{hauemmer_kalman_nlms}
C.~{Huemmer}, R.~Maas, and W.~Kellermann,
\newblock ``The {NLMS} algorithm with time-variant optimum stepsize derived
  from a {B}ayesian network perspective,''
\newblock {\em IEEE Signal Process. Lett.}, vol. 22, no. 11, pp. 1874--1878,
  2015.

\bibitem{breining_applying_nodate}
C.~Breining,
\newblock ``Applying a {Neural} {Network} for {Stepsize} {Control} in {Echo}
  {Cancellation},''
\newblock in {\em Proc. Int. Workshop Acoust. Echo Noise Control}, London, UK,
  Sept. 1997.

\bibitem{enzner_frequency-domain_2006}
G.~Enzner and P.~Vary,
\newblock ``Frequency-domain adaptive {Kalman} filter for acoustic echo control
  in hands-free telephones,''
\newblock {\em Signal Process.}, vol. 86, no. 6, pp. 1140--1156, June 2006.

\bibitem{yang_frequency-domain_2017}
F.~Yang, G.~Enzner, and J.~Yang,
\newblock ``Frequency-{Domain} {Adaptive} {Kalman} {Filter} {With} {Fast}
  {Recovery} of {Abrupt} {Echo}-{Path} {Changes},''
\newblock {\em IEEE Signal Process. Lett.}, vol. 24, no. 12, pp. 1778--1782,
  Dec. 2017.

\bibitem{malik_online_2010}
S.~Malik and G.~Enzner,
\newblock ``Online maximum-likelihood learning of time-varying dynamical models
  in block-frequency-domain,''
\newblock in {\em Int. Conf. Acoust., Speech, Signal Process.}, Dallas, USA,
  Mar. 2010, pp. 3822--3825.

\bibitem{franzen_improved_2019}
J.~Franzen and T.~Fingscheidt,
\newblock ``Improved {Measurement} {Noise} {Covariance} {Estimation} for
  {N}-channel {Feedback} {Cancellation} {Based} on the {Frequency}-{Domain}
  {Adaptive} {Kalman} {Filter},''
\newblock in {\em Int. Conf. Acoust., Speech, Signal Process.}, Brighton, UK,
  May 2019, pp. 965--969.

\bibitem{jiang_improved_2019}
T.~Jiang, R.~Liang, Q.~Wang, C.~Zou, and C.~Li,
\newblock ``An {Improved} {Practical} {State}-{Space} {FDAF} {With} {Fast}
  {Recovery} of {Abrupt} {Echo}-{Path} {Changes},''
\newblock {\em IEEE Access}, vol. 7, pp. 61353--61362, 2019.

\bibitem{kfNMF}
T.~Haubner, A.~Brendel, M.~Elminshawi, and W.~Kellermann,
\newblock ``{N}oise-{R}obust {A}daptation {C}ontrol for {S}upervised {A}coustic
  {S}ystem {I}dentification {E}xploiting a {N}oise {D}ictionary,''
\newblock in {\em Int. Conf. Acoust., Speech, Signal Process.}, Toronto, CA,
  June 2021.

\bibitem{nugraha_multichannel_2016}
A.~A. Nugraha, A.~Liutkus, and E.~Vincent,
\newblock ``Multichannel {Audio} {Source} {Separation} {With} {Deep} {Neural}
  {Networks},''
\newblock {\em IEEE Audio, Speech, and Language Process.}, vol. 24, no. 9, pp.
  1652--1664, Sept. 2016.

\bibitem{ferrara_fast_1980}
E.~Ferrara,
\newblock ``Fast implementations of {LMS} adaptive filters,''
\newblock {\em IEEE Trans. Acoust.}, vol. 28, no. 4, pp. 474--475, Aug. 1980.

\bibitem{widrow_b_adaptive_1960}
B.~Widrow and M.~E. Hoff,
\newblock ``Adaptive {Switching} {Circuits},''
\newblock in {\em {Proc.} {WESCON} {Conv.} {Rec.}}, Los Angeles, USA, Aug.
  1960, pp. 96--104.

\bibitem{shynk}
J.~J. Shynk,
\newblock ``Frequency-domain and multirate adaptive filtering,''
\newblock {\em IEEE Signal Process. Mag.}, vol. 9, no. 1, pp. 14--37, 1992.

\bibitem{7178964}
V.~{Panayotov}, G.~Chen, D.~Povey, and S.~Khudanpur,
\newblock ``Librispeech: An {ASR} corpus based on public domain audio books,''
\newblock in {\em Int. Conf. Acoust., Speech, Signal Process.}, Brisbane, AUS,
  Apr. 2015, pp. 5206--5210.

\bibitem{jeub_binaural_2009}
M.~Jeub, M.~Schäfer, and P.~Vary,
\newblock ``A binaural room impulse response database for the evaluation of
  dereverberation algorithms,''
\newblock in {\em {Int.} {Conf.} on {Digit.} {Signal} {Process.}}, Santorini,
  GR, July 2009.

\bibitem{Wen06evaluationof}
J.~Y.~C. Wen, N.~D. Gaubitch, E.~A.~P. Habets, T.~Myatt, and P.~A. Naylor,
\newblock ``Evaluation of speech dereverberation algorithms using the {MARDY}
  database,''
\newblock in {\em Proc. Int. Workshop Acoust. Echo Noise Control}, Paris, FR,
  Sept. 2006.

\bibitem{mird}
``Multi-channel impulse response database ({MIRD}),''
  \url{https://www.iks.rwth-aachen.de/en/research/tools-downloads/databases/multi-channel-impulse-response-database},
\newblock Accessed: 2020-12-04.

\bibitem{kingma2014adam}
D.~Kingma and J.~Ba,
\newblock ``{ADAM}: A method for stochastic optimization,''
\newblock {\em arXiv preprint arXiv:1412.6980}, 2014.

\end{thebibliography}

\end{document}